\documentclass[11pt,a4paper]{article}
\usepackage{jheppub}

\newcommand{\BB}{{\cal B}}

\begin{document}
\title{Measurement of \boldmath $\eta\to\pi^{0}\gamma\gamma$ branching fraction with the KLOE detector~*}

\collaboration{The KLOE-2 Collaboration}

\author[a]{D.~Babusci}
\author[b]{P.~Beltrame}
\author[c]{M.~Berlowski}
\author[a]{C.~Bloise}
\author[a]{F.~Bossi}
\author[d]{P.~Branchini}
\author[e]{B.~Cao}
\author[f,d]{F.~Ceradini}
\author[a]{P.~Ciambrone}
\author[b]{L.~Cotrozzi}
\author[g,h,1]{F.~Curciarello%
	\note{Present address: ENEA, Centro Ricerche Casaccia, Santa Maria di Galeria, Roma, Italy.}}
\author[i]{E.~Czerwi\'nski}
\author[j,k]{G.~D'Agostini}
\author[j,k,2]{R.~D'Amico}
	\note{Present address: University of Ferrara, Ferrara, Italy.}
\author[a]{E.~Dan\`e}
\author[j,k,3]{V.~De~Leo%
	\note{Present address: ENEA, Centro Ricerche Frascati, Frascati, Italy.}}
\author[a]{E.~De~Lucia}
\author[a]{A.~De~Santis}
\author[a]{P.~De~Simone}
\author[j,k]{A.~Di~Domenico}
\author[a]{E.~Diociaiuti}
\author[a]{D.~Domenici}
\author[a]{A.~D'Uffizi}
\author[j,k]{G.~Fantini}
\author[k,4]{S.~Fiore%
	\note{Present address: CERN, European Organization for Nuclear Research, Geneva, Switzerland.}}
\author[i]{A.~Gajos}
\author[i]{S.~Gamrat}
\author[j,k]{P.~Gauzzi}
\author[a]{S.~Giovannella}
\author[d]{E.~Graziani}
\author[b]{F.~Ignatov}   
\author[l]{X.~Kang}
\author[e,c]{A.~Kupsc}             
\author[m,n]{G.~Mandaglio}
\author[a,o]{M.~Martini}
\author[a]{S.~Miscetti}
\author[i]{P.~Moskal}
\author[d]{A.~Passeri}
\author[p,k]{V.~Patera}       
\author[i]{E.~Perez~del~Rio}
\author[q,v]{L.~Punzi}
\author[a,4]{F.~Sborzacchi}
\author[g,h]{M.~Schioppa}
\author[a]{A.~Selce}
\author[i]{M.~Silarski}
\author[a,r]{F.~Sirghi}
\author[s,t]{E.~P.~Solodov}
\author[u]{T.~Teubner}
\author[b,q,u]{G.~Venanzoni}
\author[b]{N.~Vestergaard}
\author[c]{W.~Wi\'slicki}
\author[e]{M.~Wolke}
\author[b]{E.~Zaid}
\affiliation[a]{Laboratori Nazionali di Frascati dell'INFN, Frascati, Italy.}
\affiliation[b]{Department of Physics, University of Liverpool, Liverpool, U.K.}
\affiliation[c]{National Centre for Nuclear Research, Warsaw, Poland.}
\affiliation[d]{INFN Sezione di Roma Tre, Roma, Italy.}
\affiliation[e]{Department of Physics and Astronomy, Uppsala University, Uppsala, Sweden.}
\affiliation[f]{Dipartimento di Matematica e Fisica dell'Universit\`a ``Roma Tre'', Roma, Italy.}
\affiliation[g]{Dipartimento di Fisica dell'Universit\`a della Calabria, Arcavacata di Rende, Italy.}
\affiliation[h]{INFN Gruppo collegato di Cosenza, Arcavacata di Rende, Italy.}
\affiliation[i]{Institute of Physics, Jagiellonian University, Cracow, Poland.}
\affiliation[j]{Dipartimento di Fisica dell'Universit\`a ``Sapienza'', Roma, Italy.}
\affiliation[k]{INFN Sezione di Roma, Roma, Italy.}
\affiliation[l]{School of Mathematics and Physics, China University of Geosciences, Wuhan, China.}
\affiliation[m]{Dipartimento di Scienze Matematiche e Informatiche, Scienze Fisiche e Scienze della Terra dell'Universit\`a di Messina, Messina, Italy.}
\affiliation[n]{INFN Sezione di Catania, Catania, Italy.}
\affiliation[o]{Dipartimento di Scienze e Tecnologie applicate, Universit\`a ``Guglielmo Marconi'', Roma, Italy.}
\affiliation[p]{Dipartimento di Scienze di Base ed Applicate per l'Ingegneria dell'Universit\`a ``Sapienza'', Roma, Italy.}
\affiliation[q]{INFN Sezione di Pisa, Pisa, Italy.}
\affiliation[r]{Horia Hulubei National Institute of Physics and Nuclear Engineering, M\v{a}gurele, Romania.}
\affiliation[s]{Budker Institute of Nuclear Physics, Novosibirsk, Russia.}
\affiliation[t]{Novosibirsk State University, Novosibirsk, Russia.}
\affiliation[u]{Department of Mathematical Sciences, University of Liverpool, Liverpool, U.K.}
\affiliation[v]{Scuola Normale Superiore, Pisa, Italy}

\dedicated{*~Dedicated to the memory of \textbf{Giuseppe Fabio Fortugno}}

\emailAdd{Marcin.Berlowski@ncbj.gov.pl}
 
\abstract{
We present a measurement of the radiative decay $\eta\to\pi^0\gamma\gamma$ using 82 million $\eta$ mesons produced in $e^+e^-\to\phi\to\eta\gamma$ process at the Frascati $\phi$-factory DA$\Phi$NE. From the data analysis $1246\pm133$ signal events are observed. By normalising the signal to the well-known $\eta\to3\pi^0$ decay the branching fraction ${\cal B}(\eta\rightarrow\pi^0\gamma\gamma)$ is measured to be $(0.98\pm 0.11_\text{stat}\pm 0.14_\text{syst})\times10^{-4}$. This result agrees with a preliminary KLOE measurement, but is twice smaller than the present world average. Results for $d\Gamma(\eta\to\pi^0\gamma\gamma)/dM^2(\gamma\gamma)$ are also presented and compared with latest theory predictions.
}
\maketitle
\flushbottom

\section{Introduction}
A long-standing interest in the double radiative $\eta\rightarrow \pi^0\gamma\gamma$ decay is due to suppression of the main contributions in the effective field theory of the strong interactions at low energies --- the Chiral Perturbation Theory (ChPT). In addition, the process is very difficult to measure because of the dominant $\eta\rightarrow 3\pi^0$ background and its branching fraction was several times revised down due to large background contamination.

The ChPT description involves tree and loop contributions from pseudoscalar mesons using an  
expansion in powers of the 4-momentum transfer, $p$, and quark masses. For the $\eta\rightarrow \pi^0\gamma\gamma$ decay, contributions from tree diagrams vanish in both ${\mathcal O}(p^2)$ and ${\mathcal O}(p^4)$ orders and therefore the first significant term is of the order ${\mathcal O}(p^6)$~\cite{Holstein:2001bt} and the predicted base line partial decay width is 0.31 eV. However, due to the cancellations of lower-order terms, the $\eta\rightarrow \pi^0\gamma\gamma$ decay width is very sensitive to details of the calculations and the model-dependent determination of the coefficients in the expansion of the chiral Lagrangian. 
Theoretical estimates of the partial decay width within ChPT are given in Refs.~\cite{Oset:2002sh} and \cite{Danilkin:2012ua}. The value for the partial decay width re-evaluated in Ref.~\cite{Oset:2008hp} is $\Gamma(\eta\rightarrow\pi^0\gamma\gamma)=0.33\pm 0.08$ eV. This value corresponds to the branching fraction ${\cal B}(\eta\rightarrow\pi^0\gamma\gamma)=(2.54\pm 0.62)\times 10^{-4}$ using the total $\eta$ decay width from the world average $\Gamma_{\eta}=1.31\pm 0.05$ keV~\cite{ParticleDataGroup:2024cfk}. 
An alternative description of the process uses the Vector-Meson Dominance (VMD) model, where one has to account for contributions given by the tree processes, where the vector mesons $\rho^0$ or $\omega$ mediate between the $\eta-\gamma$ and $\pi^0-\gamma$ vertices.
Furthermore, the latest calculations given in Ref.~\cite{Escribano:2018cwg} predict the value of ${\cal B}(\eta\rightarrow\pi^0\gamma\gamma)$ to be $(1.35\pm0.08)\times10^{-4}$ from empirical studies based on experimentally determined couplings and $(1.30\pm0.08)\times10^{-4}$ when couplings are calculated from the VMD model.

Experimentally the $\eta\rightarrow \pi^0\gamma\gamma$ decay is very hard to distinguish from background processes. The most precise measurements of ${\Gamma(\eta\rightarrow\pi^0\gamma\gamma)}$ are by the AGS/Crystal Ball \cite{Prakhov:2008zz} and the A2/Crystal Ball \cite{Nefkens:2014zdf} collaborations. The values of $\Gamma(\eta\rightarrow\pi^0\gamma\gamma)$ were determined to be $0.285 \pm 0.031_{stat} \pm 0.061_{syst}$ eV and $0.330 \pm 0.030_{tot}$ eV, respectively. The values expressed in terms of ${\cal B}(\eta\rightarrow\pi^0\gamma\gamma)$ are $(2.21\pm 0.24_{stat}\pm0.47
_{syst})\times 10^{-4}$ and $(2.56\pm 0.24_{tot})\times 10^{-4}$, respectively. Those are in tension with the preliminary result from the KLOE Collaboration  \cite{KLOE:2005hln}, ${\cal B}(\eta\rightarrow\pi^0\gamma\gamma)=(0.84 \pm 0.27_{stat} \pm 0.14_{syst})\times 10^{-4}$ which is based on a data sample of $68 \pm 23$ events. This result refers to a smaller and independent data sample than the one presented in this work.
\section{Detector}
The measurement of $\eta\rightarrow\pi^0\gamma\gamma$ was performed with the KLOE detector at DA$\Phi$NE $\phi$ factory \cite{Zobov:2007xw} in Frascati, Italy. DA$\Phi$NE is an $e^+ e^-$ collider providing collisions at 1019.5 MeV, i.e., at the $\phi$ mass, where the mesons are produced almost at rest with respect to the laboratory frame. The total production cross-section at this energy is about $3~\mu$b in the $e^+ e^-\to\phi$ process. The beams collide with a small crossing angle of about 25 mrad, resulting in the production of $\phi$ mesons with a transverse momentum of $13$ MeV/c.
The $\eta$ mesons are produced by the radiative decay $\phi\to\eta\gamma$ with a branching fraction of $\sim$1.3\%. Our measurement is based on 1.7 fb$^{-1}$ of integrated luminosity collected in 2004 and 2005.

The KLOE detector \cite{Bossi:2008aa} consists of a large (3.3 m long and 4 m diameter) cylindrical drift chamber (DC), surrounded by a finely-segmented sampling calorimeter (EMC) made of lead and scintillating fibers. The EMC consists of a cylindrical barrel and two end caps providing 98\% solid angle coverage. The whole detector is embedded in a 0.52 T magnetic field. The beam pipe at the interaction region is spherical in shape with $10$ cm radius, made of a $0.5$ mm thick beryllium-aluminium alloy.
The DC \cite{Adinolfi:2002uk} contains 12 582 drift
cells and is filled with 90\% helium--10\% isobutane low-density gas mixture. The tracks are reconstructed with the position resolutions of $\sigma_{xy}\sim150~\mu$m in the transverse plane and $\sigma_{z}\sim$2 mm. The transverse momentum resolution is $\sigma(p_t)/p_t\sim$0.4\%. Vertices are reconstructed with a spatial accuracy of $\sim$3 mm.
The EMC \cite{Adinolfi:2002zx} contains a total of 88 modules, segmented into 2440 cells with cross-section of $\sim4.4\times4.4$~cm$^2$ arranged in five layers in depth. Each cell is read out at both ends by photomultipliers. The signal amplitudes are used to determine the energy deposited in each cell, while the position along the fibres is determined by time difference. The hit cells that have both time and space information close to each other are grouped into clusters. The cluster energy $E$ is defined as the sum of the energy deposits, and the cluster time $T$ and position $R$ are the energy-weighed averages. Energy and time resolutions are $\sigma_{E}/E = 5.7\%/\sqrt{E~[\text{GeV}]}$ and $\sigma_{t}$ = 57 ps/$\sqrt{E~[\text{GeV}]} \oplus 140$ ps, respectively. Cluster positions are measured with a resolution of 1.3 cm in the coordinate transverse to the fibers and of $1.4$ cm/$\sqrt{E~[\textnormal{GeV}]}$ in the longitudinal coordinate by using timing information.
The trigger~\cite{Adinolfi:2002mvh} is based on the requirement of at least two energy deposits in the EMC above a threshold of E $>$ 50 MeV for the barrel and E $>$ 150 MeV for the end caps. The higher energy threshold required for the endcaps is chosen to reduce the machine background that contributes at small angles with respect to the beam directions. A veto on cosmic particles rejects events with high energy deposits in the outermost layers of the calorimeter and a an offline software background filter which uses calorimeter and DC information before track reconstruction rejects machine background events \cite{Ambrosino:2004qx}. The collected data are divided by an event classification filter \cite{Ambrosino:2004qx} into streams of different categories. These are stored in separate files for later analysis.

\section{Event selection}
The signature of $\phi\rightarrow\eta\gamma$, $\eta\to\pi^0\gamma\gamma$ in our measurement is five clusters not associated to tracks (neutral clusters) detected in the calorimeter with total deposited energy greater than 0.8 GeV. Since there are no charged particles in the final state we reject events with tracks connected to reconstructed clusters. In order to suppress machine background that is predominantly contributing to the signals in the end caps, the clusters must have energies at least 20~MeV and centroids at polar angles between $25^{\circ}$ and $155^{\circ}$. Moreover, a neutral cluster must satisfy the condition $| t-r/c| < min[5\sigma_{t}(E),~2 ns]$, where $t$ is the photon flight time, $r$ is the corresponding path length, $c$ is the speed of light, and $\sigma_{t}(E)$ is EMC time resolution for a cluster of given energy $E$. $2.6\times10^6$ data events with exactly five prompt clusters are selected for the analysis.

The background and signal are estimated using Monte Carlo (MC) sample generated with ten times larger luminosity than real data. The MC simulation includes the final- and initial-state radiative corrections, incorporates the machine background, and trigger efficiency on a run-by-run basis. The main physics background process is $\phi\to(\eta\to3\pi^0)\gamma$, the other considered are $\phi\to( a_0\to\pi^0\eta)\gamma$, $\phi\to (f_0\to\pi^0\pi^0)\gamma$ and $e^+e^-\to(\omega\to\pi^0\gamma)\pi^0$. Those processes are generated according to GEANFI Monte Carlo generator, based on the GEANT code \cite{Ambrosino:2004qx}, which uses PDG branching fractions and cross-sections measured by KLOE \cite{KLOE:2002kzf,KLOE:2009ehb,KLOE:2002deh,KLOE:2006vmv,KLOE:2008woc}. Moreover, the $\phi\to(\eta\to3\pi^0)\gamma$ sample is divided into four categories: a) two photons not detected, b) two photons merged into one clusters, c) one photon lost and one photon merged, d) other instances not covered by above categories. The $\eta\to\pi^0\gamma\gamma$ signal is generated according to the phase-space distribution. The preselection efficiencies evaluated with MC after the initial five photon selection are 48\% for the signal and 9\% for the sum of the backgrounds. The background yield after the preselection is given in the middle column of table~\ref{tab:final_eff2}.
\begin{table}[h!t!b!p!]
	\centering
	\begin{tabular}{ l | r | r | r }
		\hline
		Background channel                     & Generated & Preselected & Final \\
		                                       &$\times10^6$&$\times10^6$&$\times10^3$\\\hline
		$\phi\to(\eta\to3\pi^0)\gamma$ :       &&&\\
		~~~~~~~~ 2 lost                        &    -     &1.0& 29.0      \\
		~~~~~~~~ 1 lost 1 merged               &    -       &0.03& 11.9      \\
		~~~~~~~~ 2 merged                      &    -       &0.4& 1.0       \\
		~~~~~~~~ other cases                   &    -       &0.07& 3.2       \\ \hline
		$\phi\to(\eta\to3\pi^0)\gamma$ sum     &  21.8     &1.5& 45.2     \\ 
		$\phi\to (a_0\to\pi^0\eta)\gamma$      & 0.18          &0.08& 1.7     \\ 
		$\phi\to (f_0\to\pi^0\pi^0)\gamma$     & 0.45          &0.2& 0.3  \\ 
		$e^+e^-\to(\omega\to\pi^0\gamma)\pi^0$ & 0.82          &0.4& 1.6       \\ \hline
		Total contribution          & 23.2        &2.2& 48.8      \\ \hline
	\end{tabular}
	\caption{Residual backgrounds from MC. Number of events: generated (second column), after preselection (third column) and with all analysis cuts applied (last column). No events in the "generated" column for $\phi\to(\eta\to3\pi^0)\gamma$ categories because that these are determined only after reconstruction.}
	\label{tab:final_eff2}
\end{table}

\begin{figure}[!htbp]
	\begin{center}
		{\includegraphics[height=0.33\linewidth] {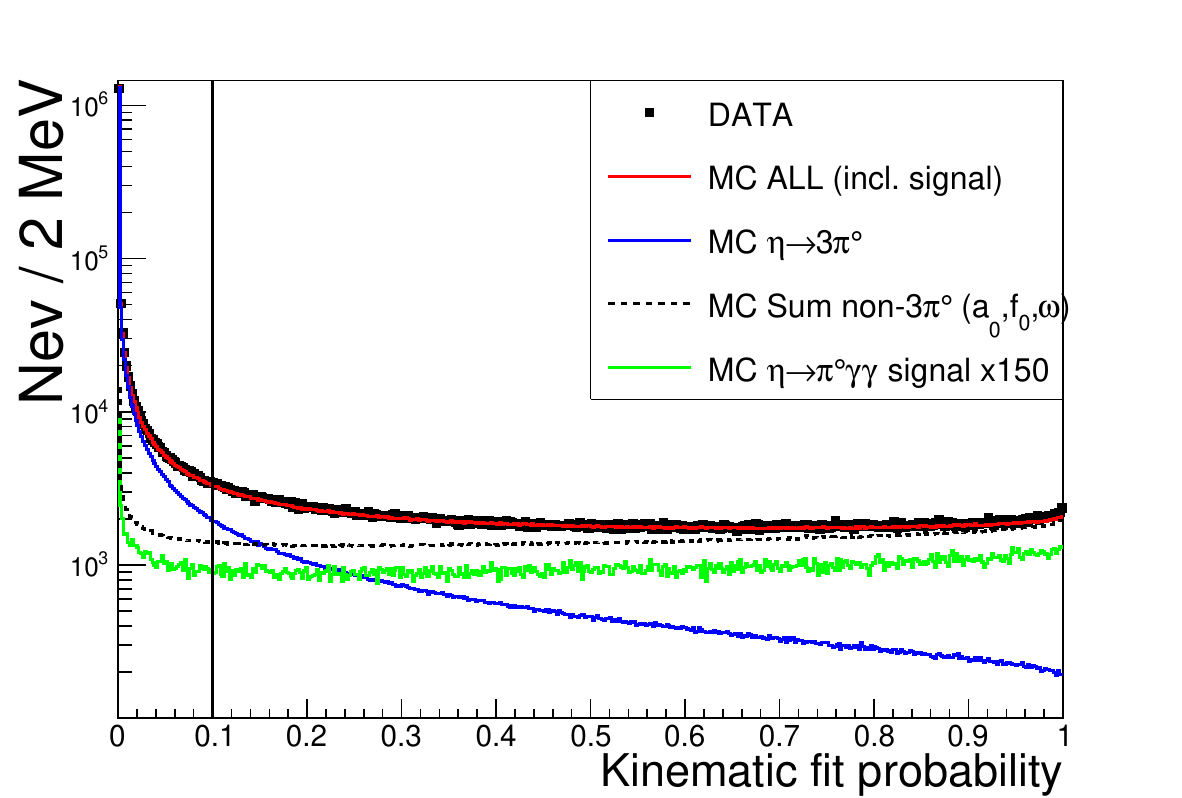}}\\
		\caption{Distribution of the 9C kinematic fit probability. Black points are data, red histogram is sum of all MC contributions represented by coloured lines: the blue line - $3\pi^0$ MC, the dotted black line - MC sum of $a_0$, $f_0$ and $\omega$, the green histogram - $\eta\to\pi^{0}\gamma\gamma$ signal MC multiplied by a factor of 150. Number of MC entries are normalised to data. The black vertical line shows the cut-off value applied.}
		\label{fig:kinfit_prob}
	\end{center}
\end{figure}

\begin{figure}[!htbp]
	\begin{center}
		{\includegraphics[height=0.25\linewidth] {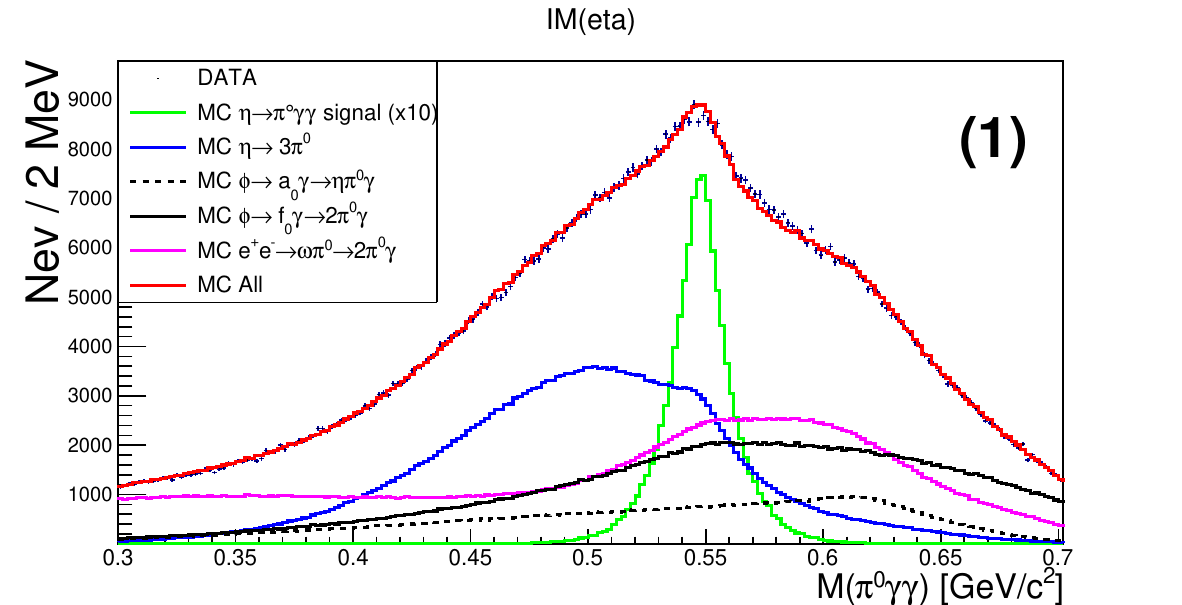}}
		{\includegraphics[height=0.25\linewidth] {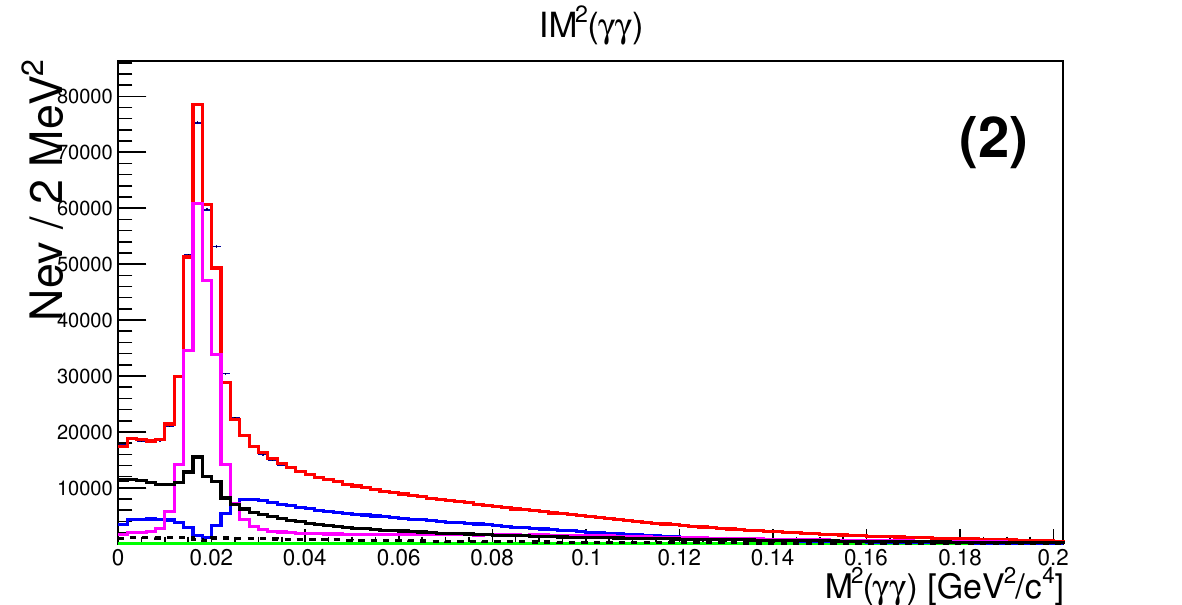}}\\
		{\includegraphics[height=0.25\linewidth] {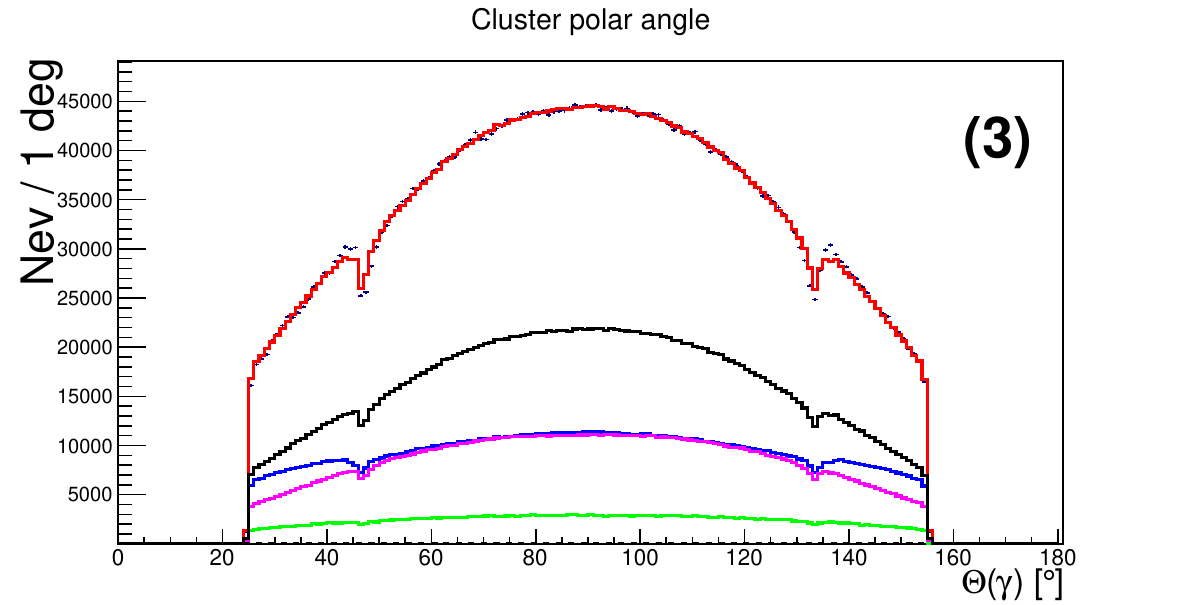}}
		{\includegraphics[height=0.25\linewidth] {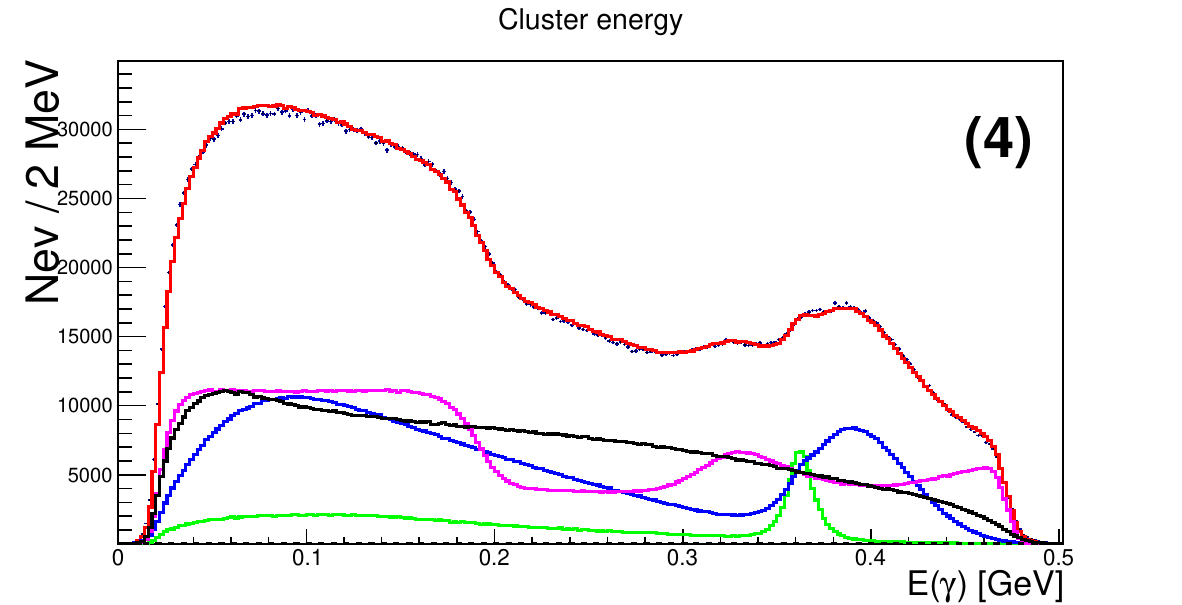}}
		\caption{Data-MC comparison after selection and cut on 9C kinematic fit probability for: (1) invariant mass of $\pi^{0}\gamma\gamma$, (2) invariant mass squared of two non-$\pi^0$ photons, (3) all 5 photon polar angles, (4) energy of each of all 5 clusters. Black points are data, red histogram is sum of all MC contributions represented by coloured lines: blue - $\eta\to3\pi^0$, black - $\phi\to (f_0\to\pi^0\pi^0)\gamma$, black dotted line - $\phi\to( a_0\to\pi^0\eta)\gamma$, magenta - $e^+e^-\to(\omega\to\pi^0\gamma)\pi^0$, green - $\eta\to\pi^{0}\gamma\gamma$ signal multiplied by a factor of 10. Number of MC entries are normalised to data. The enhancement in (4) around 360 MeV corresponds to recoil photons from the $\phi\to\eta\gamma$ decay.}
		\label{fig:control_plots}
	\end{center}
\end{figure}

\begin{figure}[!h!t!bp]
	\begin{center}
		{\includegraphics[height=0.33\linewidth]{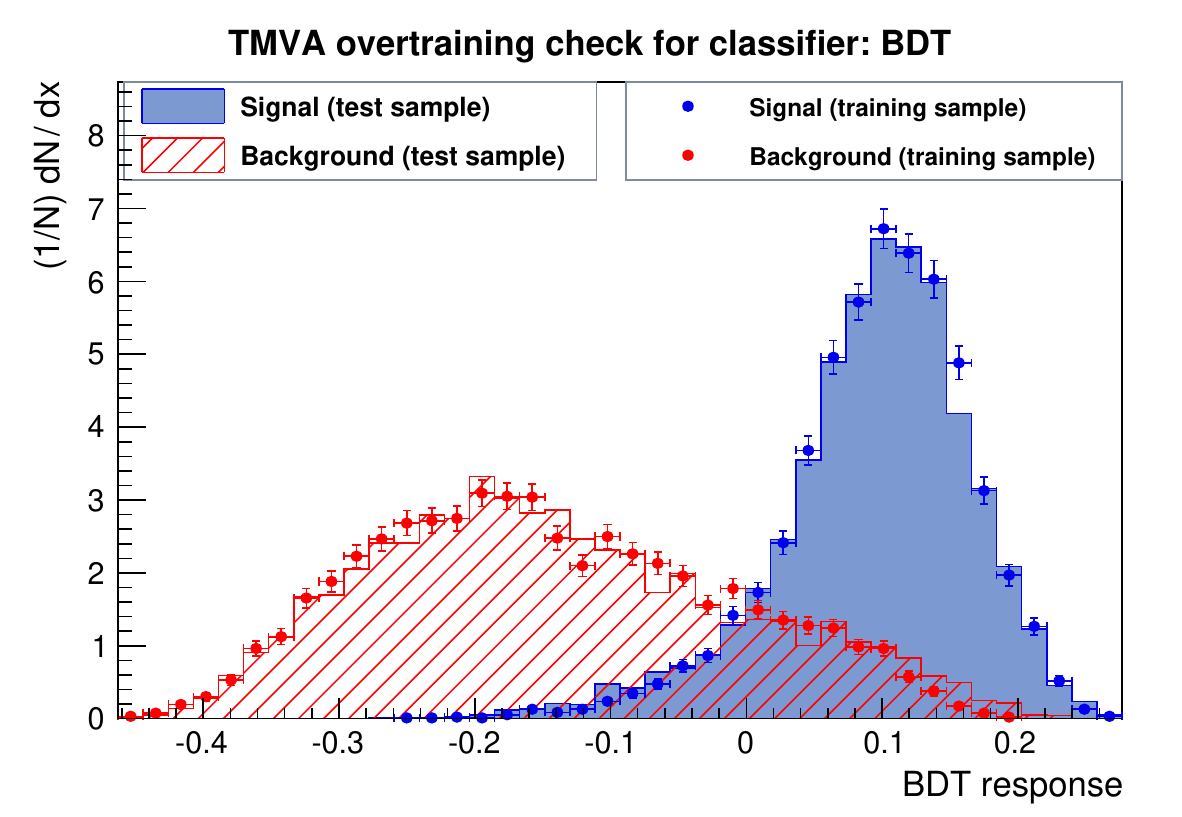}}
		{\includegraphics[height=0.33\linewidth]{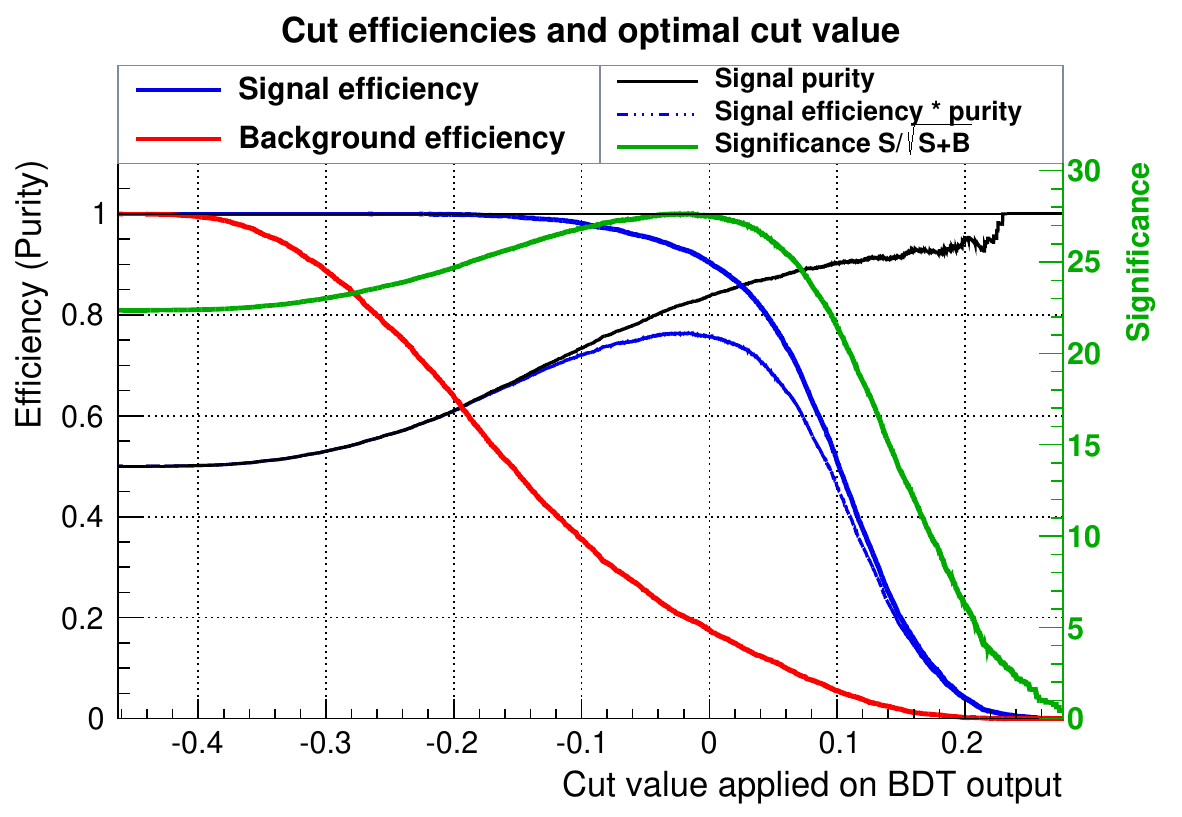}}
		\caption{(\underline{Left}) Comparison between BDT response from training (points with errors) and independent test samples (solid histograms), for EMC barrel, showing normal clusters in blue and merged clusters in red. (\underline{Right}) Signal (solid blue curve)/background (red curve) efficiency and significance (solid green) curves.}
		\label{fig:mva_overtrain}
	\end{center}
\end{figure}

A kinematic fit with 9 constraints (9C) is applied to improve the resolution of the reconstructed clusters. We use the total energy and momentum conservation and assume that all clusters originate from photons. The free parameters are the cluster positions, energies and times of flight for all five photons as well as the $e^+$ and $e^-$ beam energies and the average position of the $e^+e^-$ interaction point. The requirement of the kinematic fit probability to be greater than $10\%$ illustrated in figure~\ref{fig:kinfit_prob} removes a large fraction of events where one or more particles are undetected and where additional energy originates from the machine background. The distributions for M($\pi^{0}\gamma\gamma$) (the mass is calculated by excluding the photon with the energy closest to 363 MeV, the energy of the recoil photon from the $\phi\to\eta\gamma$ decay), M$^2(\gamma\gamma)$ (mass squared of two non-$\pi^0$ photons, where $\pi^0$ is defined as two photons closes to its mass), all 5 clusters cluster polar angle and energy after the cut on kinematic fit probability are presented in figure~\ref{fig:control_plots}.
\\
\\
The background from $\phi\to (a_0\to\pi^0\eta)\gamma\to5\gamma$ production is reduced by using the 11C kinematic fit with additional constraints, where after excluding the photon with the energy closest to 363 MeV, the energy of the recoil photon from the $\phi\to\eta\gamma$ decay, one pair of the photons is required to originate from $\eta$ meson and the other from $\pi^0$ meson. From all possible combinations we keep the one with the lowest $\chi^2$ and discard the event if the probability of the fit exceeds $10\%$. A similar procedure is used to suppress processes with $\pi^0\pi^0$ in the final state. Here the additional constraints require two pairs of photons from $\pi^0$ mesons. These events are also discarded if the fit probability is greater than $10\%$.
To further reduce the backgrounds with neutral pion pairs in the final state we use the $\chi^2$ defined as follows:
\begin{equation}
	\chi^2(2\pi^{0})=\frac{[M(\gamma_{1}\gamma_{2})-m(\pi^{0})]^{2}}{\sigma^{2}(\gamma_1\gamma_2)}
	+\frac{[M(\gamma_{3}\gamma_{4})-m(\pi^{0})]^{2}}{\sigma^{2}(\gamma_3\gamma_4)},
	\label{pchi2_1}
\end{equation}
where $M(\gamma\gamma)$ is the invariant mass of a photon pair, $m(\pi^{0})$ is the mass of neutral pion and $\sigma(\gamma\gamma)$ is the uncertainty on the $\gamma\gamma$ invariant mass. In our case the angular uncertainty is negligible in respect to the energy one and therefore the uncertainty of the invariant mass $\sigma(\gamma\gamma)$ can be expressed as:
\begin{equation}
	\frac{\sigma(\gamma_i\gamma_j)}{M(\gamma_{i}\gamma_{j})}=\frac{1}{2}\sqrt{\frac{\sigma^{2}(E_i)}{E_{i}^{2}}+\frac{\sigma^{2}(E_j)}{E_{j}^{2}}}\ ,
	\label{pchi2_2}
\end{equation}
where $E_{i}$ and $E_{j}$ are cluster energies and $\sigma(E_{i})$ and $\sigma(E_{j})$ their corresponding energy resolutions. We require the $\chi^2(2\pi^{0})$ value for the best combination to be above 2.6. This value is optimised to reduce the contribution of events with two $\pi^{0}$ by $90\%$ and to preserve $85\%$ of the signal.\\
In order to suppress the background originating from $\eta\rightarrow3\pi^0$ with two lost photons we construct a kinematic fit which, in addition to the standard 9C constraints, takes into account properties of events where two photons are lost. These events are constrained by the total energy deposited in the EMC and the missing momentum. The hypothesis for the fit is $\phi\to\gamma(\eta\to3\pi^0)$ decay with energy and momentum conservation.\\
The first step of the fit procedure is to find the initial values for the momenta of the two lost photons. We assume that the missing four-momentum is the sum of the four-momenta of the two missing photons and each pair of the reconstructed photons originate from different $\pi^{0}$ decay. The cases when a whole $\pi^{0}$ meson is lost are covered by the 11C kinematic fit with 2$\pi^{0}$ mass constrain. This provides six constraints needed to determine the two unmeasured momenta. A solution only exists if the invariant mass constructed with the missing four-momentum and the four-momenta of the two candidate partners of the missing photons is greater than $m(\pi^0)$. If there are no solutions, the event is considered as one without two lost photons and is used for further analysis. If at least one real solution is found, it is used as a starting point for the iterative kinematic fit. The procedure reduces the $\eta\rightarrow3\pi^0$ with two lost photons by 15\% while keeping more than $97\%$ of the signal.\\
For the $\eta\rightarrow3\pi^0$ case with merged photons, when two photons are sharing the same cluster, one of the Multivariate Data Analysis (MVA) methods, Boosted Decision Trees (BDT)~\cite{Hocker:2007ht,Voss:2007jxm} is used. The method is based on the properties of the clusters as well as their energy and cell multiplicity to distinguish between the "normal" and the "merged" clusters. The BDT is trained using the signal $\eta\to\pi^0\gamma\gamma$ MC events and the $\eta\rightarrow3\pi^0$ MC events with only merged clusters (five reconstructed clusters). Since the calorimeter has different structure in the end caps and barrel part, the BDT training is split into two separate parts according to the position of the cluster. The clusters consisting only of a single cell are not considered.
The training samples for both categories were 5000 clusters.
The test phase is carried out on the 5000 independent clusters for the normal/merged class in barrel/end caps. The cut-off value for the BDT response (-0.0365 for the barrel
and -0.0136 for end caps) is chosen to maximise the significance, defined as ${S}/{\sqrt{S+B}}$, where S is the signal (the normal clusters), B is the background (the merged clusters). The single cluster efficiencies are 96.6\% for the normal class and 20.3\% for the merged in the barrel and 91.1\% and 18.4\%, respectively, in the end caps. The BDT response is checked for all clusters in the event. If even one of the clusters is marked as merged by the BDT, the entire event is removed from the analysis.
The test results for the barrel part of the calorimeter clusters are shown in figure~\ref{fig:mva_overtrain}.
The last column in table~\ref{tab:final_eff2} shows the number of events generated, remaining after preselection and surviving all the selection cuts.\\
To summarize: we select five prompt photons with total energy sum greater than 800~MeV. Each photon must be emitted with polar angle between $25^{\circ}$ and $155^{\circ}$ and energy above 20~MeV. Photon energies and times are corrected using 9C kinematic fit and the event is retained applying a cut value at 10\% probability. The backgrounds from $\phi\to( a_0\to\pi^0\eta)\gamma$, $\phi\to (f_0\to\pi^0\pi^0)\gamma$ and $e^+e^-\to(\omega\to\pi^0\gamma)\pi^0$ are reduced using two 11C kinematic fits with probability cut at 10\%, neutral pion pairs are further suppressed by using $\chi^2$ cut at value of 2.6 and merged clusters are treated by BDT method. The number of selected data events is 45698 and the final signal efficiency evaluated with MC simulation is $\epsilon_{S}=(15.45\pm 0.09)\%$.
\section{The normalisation channel}
\begin{figure}[!htbp]
	\begin{center}
		{\includegraphics[height=0.32\linewidth]{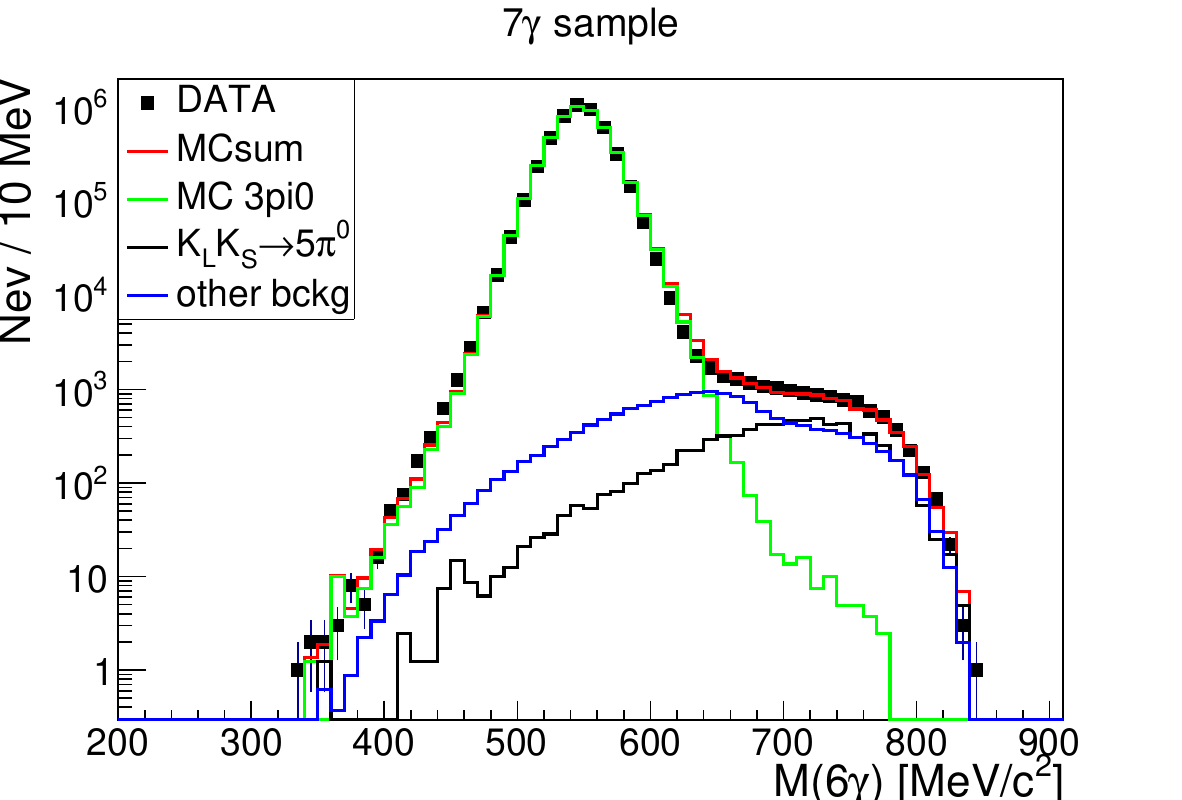}}
		{\includegraphics[height=0.32\linewidth]{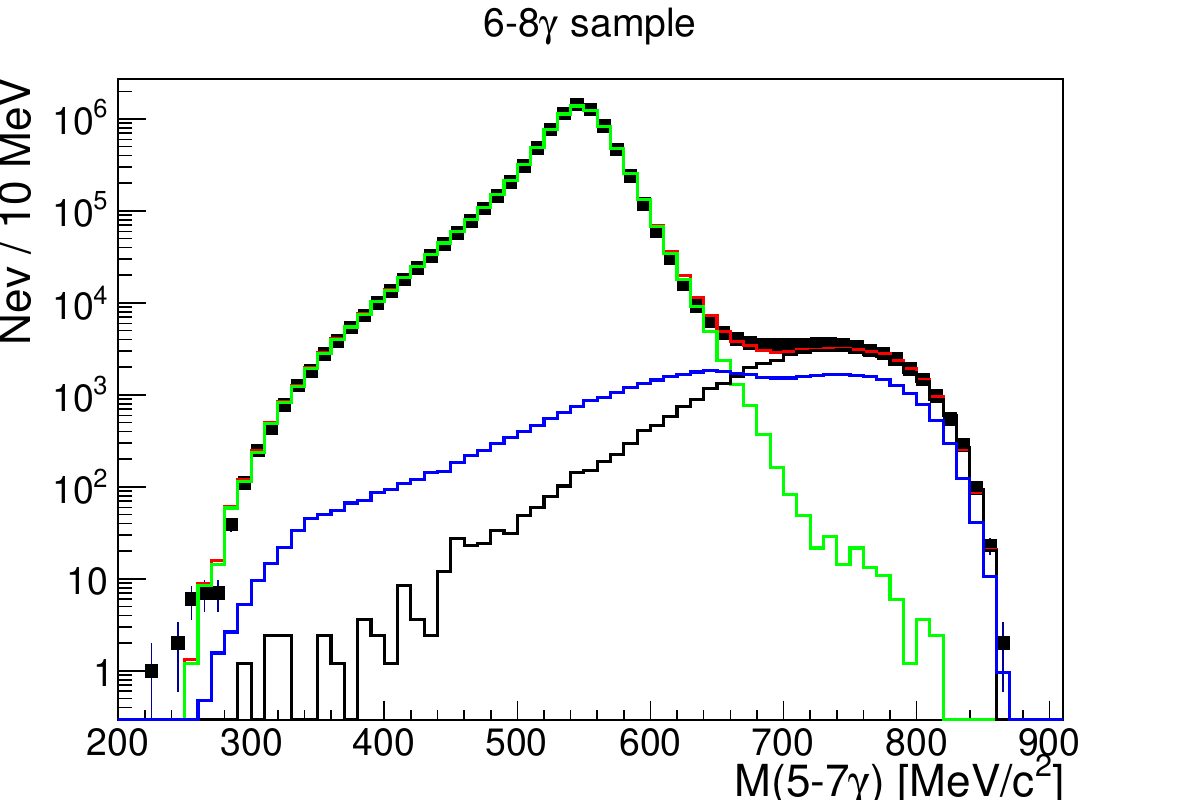}}
		\caption{Distributions of $M(6\gamma)$ (\underline{left}) and  M(5--7$\gamma)$ (\underline{right}) in log scale for the $Y$ axis after the fit. Data as black points, red histogram is sum of all MC contributions: green histogram - $\eta\to3\pi^0$ channel, black histogram - $K_S K_L\rightarrow5\pi^0$ background, blue histogram - all other backgrounds.}
		\label{fig:3pi0_im}
	\end{center}
\end{figure}

The decay $\eta\to3\pi^0$ is chosen as the normalisation channel since it has similar topology and the well known branching fraction value of $\BB(\eta\to3\pi^0)=\BB_{3\pi^0}=(32.57\pm0.21)\%$ \cite{ParticleDataGroup:2024cfk}. The selection criteria for the photons are the same as for the $\eta\to\pi^0\gamma\gamma$ signal extraction: the minimum photon energies, their polar angles, the total energy deposited and time windows. To reject events with photon splitting (i.e. when a single photon produces two separate clusters) probability we require that each pair of clusters has an opening angle greater than $18^{\circ}$. As for the signal the resolution in energy and time is improved using a 11C kinematic fit with total energy-momentum conservation and time of flight of photons constraint. The probability value of the kinematic fit $>10\%$ is required to suppress poorly reconstructed events. We exclude the recoil photon (the one closest to 363 MeV) and require the invariant mass of the six remaining photons to be consistent with the $\eta$ meson mass. The efficiency for the $3\pi^0$ signal is $\epsilon_{3\pi^0}= (17.94\pm0.01)\%$ with backgrounds well below the $1\%$ level. We use an unbinned 3-component maximum likelihood fit to the $M(3\pi^0)$ mass spectrum, where the $\eta\to3\pi^0$ and background shapes are from the MC simulation. The number of events for the subsamples consisting of 7 or the sum of 6, 7 or 8 photons are presented in table~\ref{tab:3pi0_numbers}. For 6$\gamma$ and 8$\gamma$ samples, $M(3\pi^0)$ is constructed using 5 and 7 photons, respectively. The main remaining backgrounds after the selection are $\phi\to K_S K_L \to (2\pi^0)(3\pi^0)$ and $\phi\to\eta'(\to \eta\pi^0\pi^0)\gamma$ with subsequent decay $\eta\to\gamma\gamma$.
\begin{table}[h!t!b!p!]
	\centering
	\begin{tabular}{  l | c  c  c  c  c }
		\hline
     Sample & $N_{data}$ & $N_{3\pi^0}$ & $N_\text{bckg}$ & $\epsilon_{3\pi^0}$(\%) &  $N_\text{norm}\times10^6$\\ \hline
 7$\gamma$ & 4804600 & 4790792 & 13800 &  $17.94\pm0.01$ & $81.98\pm0.53$\\
6, 7 or 8$\gamma$ & 7124843 & 7078774 & 46062 &  $27.12\pm0.01$ & $80.15\pm0.52$\\
 \hline
	\end{tabular}
\caption{Number of events for $\eta\to3\pi^0$ selection, number of background events, $\eta\to3\pi^0$ efficiency and the corresponding number of $\eta$ mesons produced for the two samples: $7\gamma$ and 6, 7 or $8\gamma$ after the fit.}
\label{tab:3pi0_numbers}
\end{table}
The distributions of the invariant mass of 6 and 5--7 photons are presented in figure~\ref{fig:3pi0_im} for the 7 subsample and for the sum of the 6-, 7- and 8-photon subsamples.
The corresponding number of $\eta$ mesons produced ($N_\text{norm}$) is calculated using the 7 photon sample. For the systematic uncertainty we consider three sources: cluster merging, photon splitting and analysis cut on kinematic fit probability. The first two systematic uncertainties are estimated using 6-, 7- and 8-photon subsample by taking the half of a difference from 7 photon $N_\text{norm}$ measurement, which is 1.1\%. The uncertainty associated with the kinematic fit probability cut is studied by varying its value within 5\% from the nominal value, which gives the systematic uncertainty of 0.6\%. The total 1.3\% systematic uncertainty is given by the quadratic sum of the individual contributions, assuming that all the sources are independent. The calculated final number of $\eta$ mesons is:  $N_\text{norm}=N_{3\pi^0}/\epsilon_{3\pi^0}/\BB_{3\pi^0}=(81.98\pm0.53_\text{stat}\pm1.03_\text{syst})\times10^6$.

\section{The signal extraction}
\begin{figure}[!htbp]
	\begin{center}
		{\includegraphics[height=0.32\linewidth]{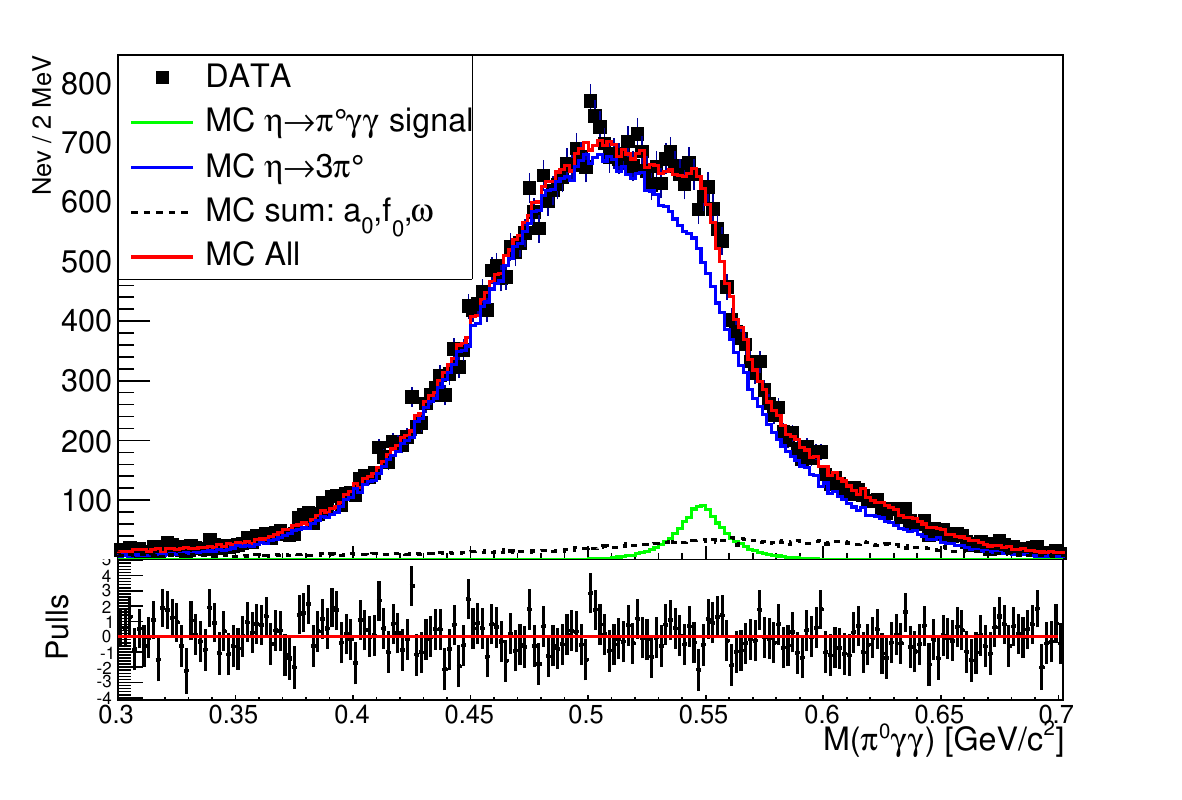}}
		{\includegraphics[height=0.32\linewidth]{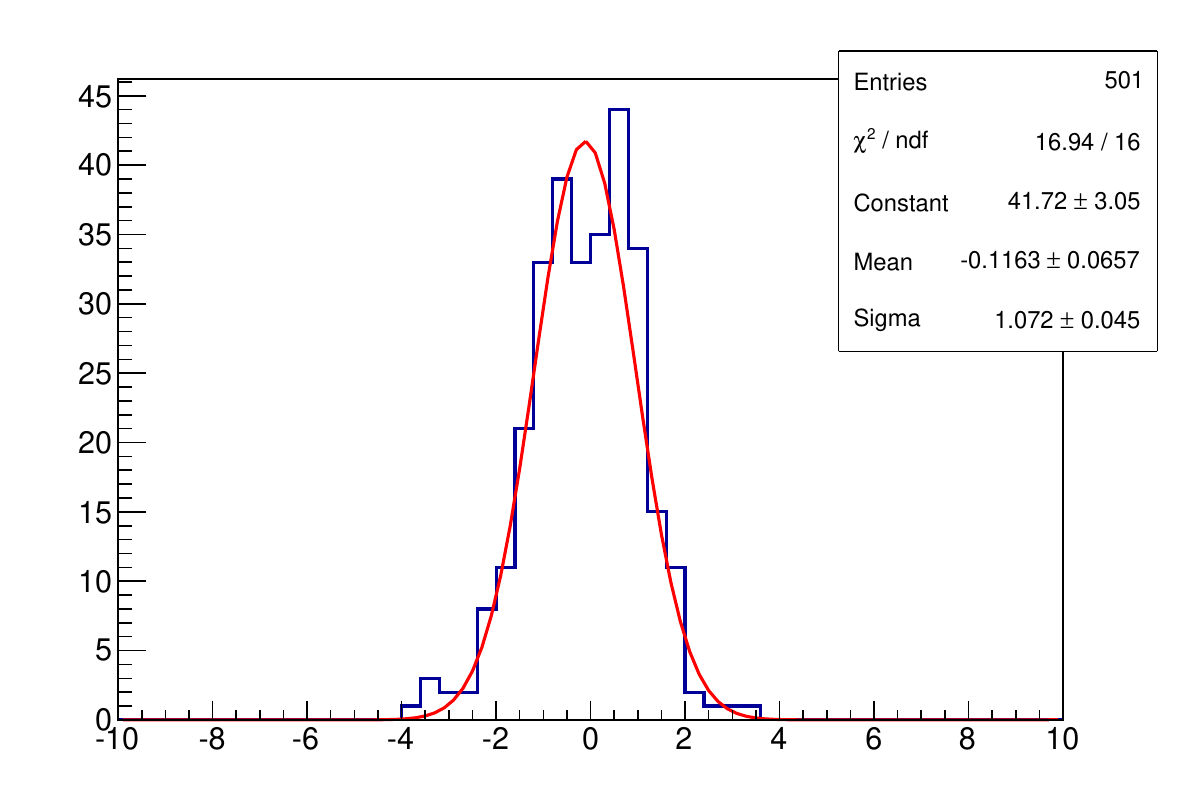}}\\
		\caption{(\underline{Left}) Result of the fit to the $\pi^0\gamma\gamma$ invariant mass distribution. Data are presented as black points, red histogram is the sum of all MC contributions, the green line represents the $\eta\to\pi^0\gamma\gamma$ signal, the blue line shows the $3\pi^0$ background and the dotted black line is the sum of the $a_0$, $f_0$ and $\omega$ channels. The residuals are presented below the distribution. (\underline{Right}) Distribution of residuals with Gaussian fit superimposed (red line), the fitted Gauss parameters are in the box.}
		\label{fig:final_imeta}
	\end{center}
\end{figure}

To determine the number of the signal events we use an unbinned 3-component maximum likelihood fit to the $M(\pi^0\gamma\gamma)$ mass spectrum. In the fit, the signal and background shapes are from the MC simulation and all non-peaking backgrounds from $a_0$, $f_0$ and $\omega$ decays are summed. In the MC $\phi\to\eta\gamma$ is generated according to its known $3~\mu b$ production cross-section while for the $a_0$, $f_0$ and $\omega$ production cross-sections and decay ratios are taken from the corresponding KLOE results \cite{KLOE:2002kzf,KLOE:2009ehb,KLOE:2002deh,KLOE:2006vmv,KLOE:2008woc}.
The invariant mass $M(\pi^0\gamma\gamma)$ is calculated by excluding the photon with the energy closest to 363 MeV, the energy of the recoil photon from the $\phi\to\eta\gamma$ decay. The $\chi^2$-test of the fit result gives $\chi^2/\text{n.d.f.} = 215/200$ corresponding to the p-value of 22\%. The results of the fit are presented in figure~\ref{fig:final_imeta}.
The numbers of events for each component from the fit are given in table~\ref{tab:final_imeta} including the statistical uncertainties and  
 \begin{table}[h]
 \centering
	\begin{tabular}{  l | r |cc}
		\hline
  &&\multicolumn{2}{|c}{Correlation coefficients}\\
		Fit component                             & Nb of events & $3\pi^0$  background &  Non-$3\pi^0$ background  \\ 
	\hline
  $\pi^0\gamma\gamma$                 & $ 1246 \pm 133 $ &$-0.40$& $-0.02$ \\ 
		$3\pi^0$  background                      & $ 45232 \pm 307$  &&$-0.52$ \\ 
		Non-$3\pi^0$ background                   & $ 3546 \pm 190$  &&  \\ \hline
	\end{tabular}
	\\ \caption{The 3-component fit results including correlation coefficients.}
	\label{tab:final_imeta}
\end{table}
the correlation coefficients.\\
The branching fraction ${\cal B}(\eta\to\pi^0\gamma\gamma)$ is calculated as ${\cal B}(\eta\to\pi^0\gamma\gamma)=N_S/\epsilon_S/N_\text{norm}=N_S/N_{3\pi^0}\cdot\epsilon_{3\pi^0}/\epsilon_S\cdot\BB_{3\pi^0}$, where $N_S$ is the number of the signal events from the fit and $\epsilon_S$  is the signal efficiency from the MC simulation, $N_\text{norm}$ is the final number of $\eta$ mesons. The result is $(0.98\pm0.11)\times 10^{-4}$, where the uncertainty is statistical only.

\begin{figure}[!htbp]
	\begin{center}
		\includegraphics[width=0.6\textwidth]{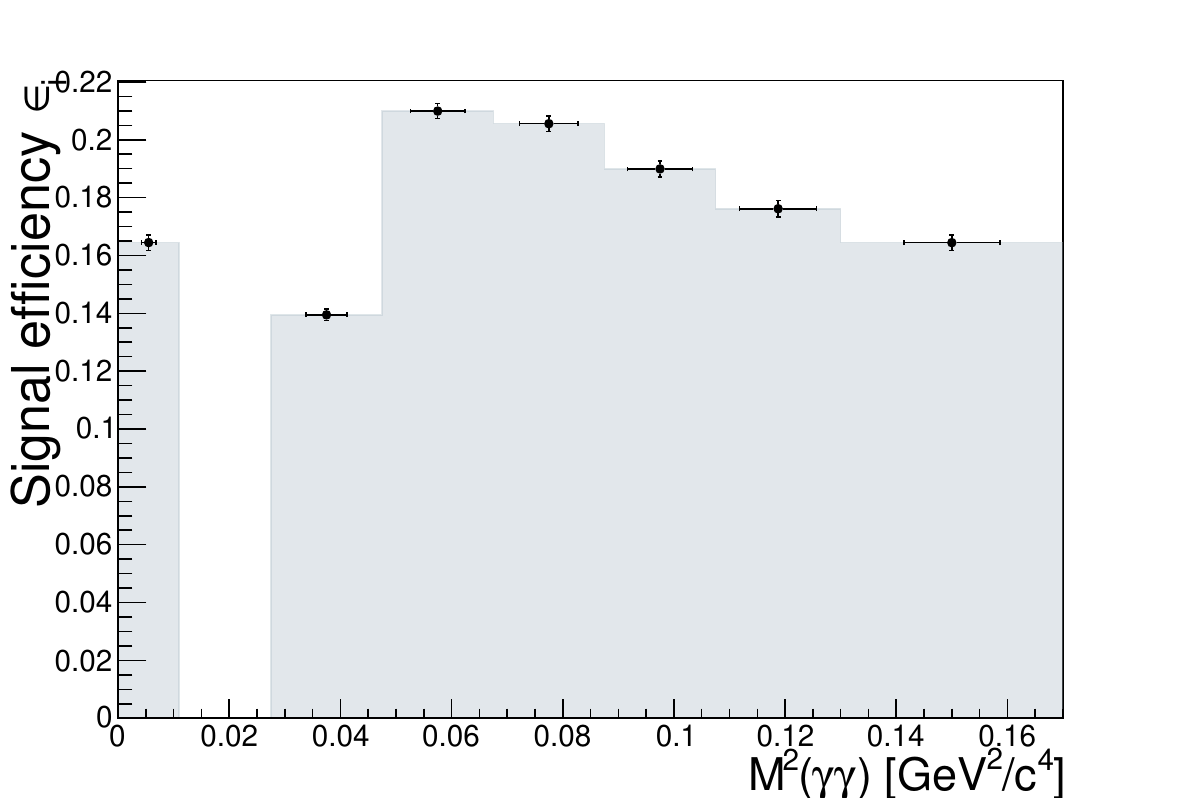}
		\caption{ Detection efficiencies $\epsilon_i$ for $\eta\to\pi^0\gamma\gamma$ in bins of $M^2(\gamma\gamma)$, evaluated from the MC simulation. The resolution in $M^2(\gamma\gamma)$ is indicated as horizontal $\pm 1\sigma$ lines at the bin centres. The bin boundaries are represented by the histogram bin widths. The region $(0.0110,0.0275)$ GeV$^2$/c$^4$ is excluded due to low efficiency of the $\pi^0\pi^0$ veto.}
		\label{fig:dg_eff}
	\end{center}
\end{figure}

\begin{figure}[!htbp]
	\begin{center}
		{\includegraphics[height=0.5\linewidth,trim={0 0 0cm 0cm},clip]{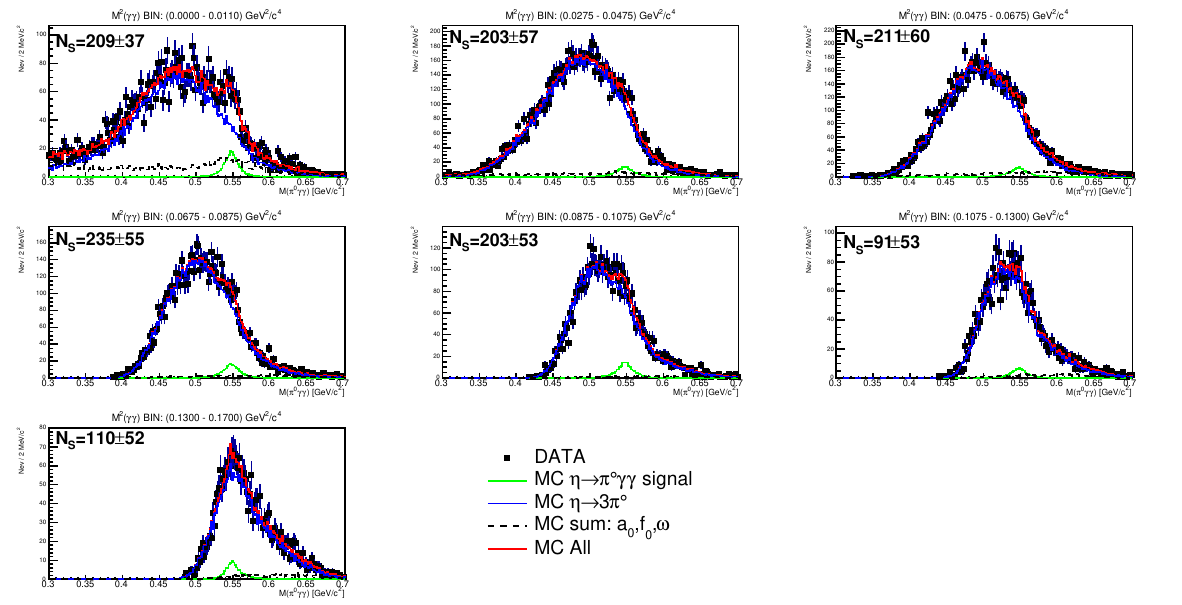}}
		\caption{ Fits to the $\pi^0\gamma\gamma$ invariant mass distribution as in figure~\ref{fig:final_imeta}(Left) for each of the $M^2(\gamma\gamma)$ bins. Text in upper-left corner shows the number of the $\eta\to\pi^0\gamma\gamma$ signal events extracted from the fits. The bin ranges are presented above the distributions.}
		\label{fig:dg_fits}
	\end{center}
\end{figure}

To measure the decay rate as a function of the invariant mass squared $M^2(\gamma\gamma)$ of the two bachelor photons (not originating from the $\pi^0$ decay) the data are divided into seven $M^2(\gamma\gamma)$ bins. The bin ranges are $1^\text{st}:(0,0.011)$, $2^\text{nd}:(0.0275,0.0475)$, $3^\text{rd}:(0.0475,0.0675)$, $4^\text{th}:(0.0675,0.0875)$, $5^\text{th}:(0.0875,0.1075)$, $6^\text{th}:(0.1075,0.13)$, $7^\text{th}:(0.13,0.17)$ in GeV$^2$/c$^4$ units. The ranges are illustrated in figure~\ref{fig:dg_eff} by the histogram of the detection efficiency in each bin. The region close to the neutral pion mass squared $(0.0110,0.0275)$ GeV$^2$/c$^4$ is absent since the events are removed by the veto on the $\pi^0\pi^0$ pairs.
The corresponding bin efficiencies shown in figure~\ref{fig:dg_eff} are obtained from the signal MC with constant matrix element for $\eta\to\pi^0\gamma\gamma$. The $i$-th bin efficiency $\epsilon_i$ is defined as the ratio of the events that are reconstructed in the bin to the generated with $M^2(\gamma\gamma)$ within the bin limits.   
The number of events in each $M^2(\gamma\gamma)$ bin is determined using the same fitting procedure as for the branching fraction determination, repeated for each bin. The results are presented in figure~\ref{fig:dg_fits}.
The value of $d\Gamma(\eta\to\pi^0\gamma\gamma)/dM^2(\gamma\gamma)$ is then calculated according to the formula:
\begin{equation}
	\left.\frac{d\Gamma}{dM^2}\right|_i=\frac{N_S^{(i)}}{\epsilon_S^{(i)} \cdot N_\text{norm}}\frac{\Gamma}{w_i}\ ,
	\label{eq:dgdm2}
\end{equation}
where $N_S^{(i)}$ is signal yield, $w_{i}$ is width of the $i$-th bin, $\epsilon_S^{(i)}$ is the efficiency for signal for a given bin, $N_\text{norm}$ is the final number of $\eta$ mesons, and $\Gamma=(1.31\pm0.05)$ keV is the PDG full width of the $\eta$ meson. 
The statistical uncertainties correspond to the uncertainties of $N_S^{(i)}$ from the fit. The $N_S^{(i)}$ sum is $1262 \pm 140$, in good agreement with the total signal yield $1246 \pm 133$.
\section{Systematic studies}\label{sec:syst}
The list of systematic effects that are considered includes the trigger conditions, background filter, data streaming, consistency between data acquisition periods in the analysis sample, photon counting efficiency, identification of the recoil photon, selection cuts applied in the analysis, biases in the fitting procedure, normalisation error, signal MC model and finally the effect of a data-MC difference observed for small values of $M(\gamma\gamma)$ region (below 25 MeV). For $d\Gamma/dM^2$, additional uncertainty due to incorrect assignment of photons as originating from the $\pi^0$ decay is also considered. The systematic uncertainties are summarised in table~\ref{tab:syst}.

The systematic effect due to the trigger conditions is investigated using the MC simulated signal sample and found to be negligible. To estimate the systematic uncertainties due to the background filter and data streaming procedure we use a data sample collected without streaming or background filtering conditions, scaled down by a factor of ten. The determined systematic uncertainty is estimated to be 1\% for each effect. However since in our measurement we are normalising to a channel which was collected with exactly the same trigger during the same time period, both signal and $\eta\to3\pi^0$ are in the same data stream under the same background filter, the effects of trigger, background filter and data streaming mostly cancel out in the branching fraction and $d\Gamma/dM^2$ measurement.

Since the sample comes from two separate data acquisition periods we have checked that the results obtained separately on each sample are consistent within the errors.

The systematics from the event selection is studied by varying separately each of the selection cuts within its own resolution. The resulting half of the difference between the maximum and the minimum results separately for each selection cut is assumed as an estimate of the systematic uncertainty. The total uncertainty is given by the quadratic sum of the individual contributions, assuming that all the sources are independent.

In order to estimate the systematic uncertainty due to the pairing procedures used in the analysis the $\eta\to\pi^0\gamma\gamma$ is studied with MC generated signal sample by matching the generated photon pairs to the selected $\pi^0$ candidates and to the unpaired photons. The fraction of events with incorrectly combined photons is 5.1\% according to MC studies. Alternative fit is performed by removing the wrongly combined events from MC signal sample. The change in the branching fraction value is 0.6\% and it is found to be negligible. The difference between the two fit results for $d\Gamma/dM^2$ is considered as a source of systematic uncertainty.

The uncertainties associated with the analysis conditions is studied by varying the cut values sequentially using their own resolution for each of them. Since all kinematic fits are correlated we vary the probability cut separately for each of them and take the maximum and the minimum changes in the branching fraction value and $d\Gamma/dM^2$ with respect to the nominal results for all of them. The resulting half of the difference between the maximum and the minimum results separately for each selection, is considered as the systematic uncertainty.

Different binning schemes of the $M(\pi^0\gamma\gamma)$ distribution and ranges for the fitting procedure are used to estimate the associated uncertainties. The changes in the branching fraction value and $d\Gamma/dM^2$ with respect to the nominal results are found to be negligible. Resolution effects on the $d\Gamma/dM^2$ distribution have been studied with the MC simulation resulting in a negligible modification of the distribution in the chosen binning scheme.

The systematic uncertainty on the normalisation is estimated to be 1.3\%
as explained in section 4.

To estimate the systematic uncertainty from MC model, the phase space distribution for $M^2(\gamma\gamma)$ in $\eta\to\pi^0\gamma\gamma$ is replaced by the model proposed in Ref.~\cite{Escribano:2018cwg}. The branching fraction and $d\Gamma/dM^2$ values obtained using it are consistent with the nominal analysis results.

For the $d\Gamma/dM^2$ studies we distinguish between the normalisation uncertainty common to all points and uncorrelated ones. Here for each of the statistical or the systematic uncertainty we subtract in quadrature from each the $d\Gamma/dM^2$ bin value the corresponding uncertainty obtained from the branching fraction analysis.

\begin{figure}[!htbp]
	\begin{center}
		{\includegraphics[height=0.35\linewidth]{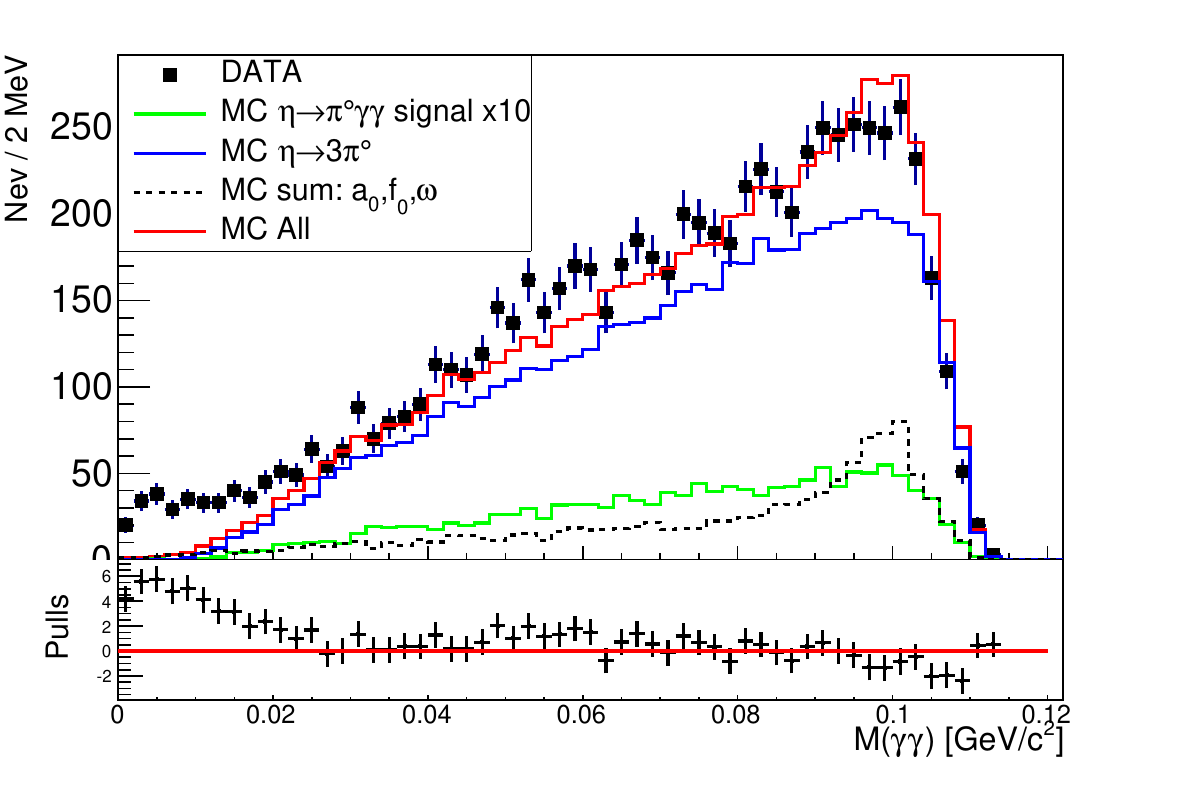}}
		\caption{Small invariant mass region of non-$\pi^0$ photons. Black points are data, red histogram is sum of all MC contributions: green - $\eta\to\pi^0\gamma\gamma$ signal (upscaled be a factor of 10), blue - $\eta\to3\pi^0\to5\gamma$, and dashed black line corresponds to the sum of the $a_0$, $f_0$ and $\omega$ channels. All MC lines are normalised according to the results of the 3-component fit to $M(\pi^0\gamma\gamma)$. The residuals are presented below the distribution.}
		\label{fig:imgg_small}
	\end{center}
\end{figure}

We observe a difference between MC and data for the $M(\gamma\gamma)$ values below 25 MeV, see figure~\ref{fig:imgg_small}. Since both clusters are close in space and time and one of them has much smaller energy we can associate them with photon splitting events originating from $\phi\to\gamma\eta\to3\gamma$, $\phi\to\gamma\pi^0\to3\gamma$ or $e^+e^-\to\gamma\gamma$ processes. The effect is investigated by applying cuts removing events with $M(\gamma\gamma)$ less then 20, 25, 30 and 50 MeV.
The extracted $d\Gamma/dM^2$ value in the standard analysis for the first bin is $1.96\pm0.37_\text{stat}$ eV/(GeV$^2$/c$^4$). By removing the events with small $M(\gamma\gamma)$ the largest change is $\Delta=0.11$ eV/(GeV$^2$/c$^4$) for $M(\gamma\gamma)>25$ MeV, which corresponds to a negligible contribution compared to other systematic effects. In addition, the uncertainty of $\Delta$, $\sigma(\Delta)$, is calculated taking into account the correlation between the samples \cite{Barlow:2002yb}. The ratio $\Delta / \sigma(\Delta)$ = 1.8 confirms that the change can be interpreted as no significant systematic effect (see Ref.~\cite{Barlow:2002yb}) on the value of $d\Gamma/dM^2$ in the first bin. The same checks are performed for the $\BB(\eta\to\pi^0\gamma\gamma)$ value and the effect is also negligible.

The systematic uncertainties are summarised in table~\ref{tab:syst}.
\begin{table}[h!t!b!p!]
	\centering
	\begin{tabular}{ l | r | c c c c c c c}
		\hline 
		 &\multicolumn{8}{|c}{Relative uncertainties in \%}\\ \cline{2-9}
		\multicolumn{1}{c|}{Source of} &${\cal B}(\eta\rightarrow\pi^0\gamma\gamma)$& \multicolumn{7}{|c}{$d\Gamma/dM^2$ bin number }\\
	  \multicolumn{1}{c|}{uncertainty}&& $1^\text{st}$ & $2^\text{nd}$ & $3^\text{rd}$ & $4^\text{th}$ & $5^\text{th}$ & $6^\text{th}$ & $7^\text{th}$\\ \hline
		Preselection & 11 & 38	&11	&17	&19	&1	&56	&34		\\
		Analysis cuts & 9 & 27	&42	&25	&21	&10	&46	&41	\\
        Photon pairing & 0.6 & 2 & 4 & 1 & 1 & 4 & 2 & 8\\
		Normalisation & 1 &  1 & 1 & 1 & 1 & 1 & 1 & 1 \\
		\hline
		Total & 14 & 46	& 44 & 30 & 28 & 10 & 73 & 54		\\ \hline
	\end{tabular}
\caption{The relative systematic uncertainties of the $\BB(\eta\rightarrow\pi^0\gamma\gamma$) and all bins of the $d\Gamma(\eta\to\pi^0\gamma\gamma)/dM^2(\gamma\gamma)$ (all values are given in \%). 
The effects of fitting bias, MC model, small $M(\gamma\gamma)$ and run period result in a negligible contribution to the systematic uncertainty.}
	\label{tab:syst}
\end{table}
\section{Results}
Using a data sample of $1.7\text{ fb}^{-1}$, corresponding to $(81.98\pm0.53_\text{stat}\pm1.03_\text{syst})\times10^6$ $\eta$ mesons collected in the KLOE detector, $1246 \pm 133$ $\eta\rightarrow\pi^0\gamma\gamma$ events are selected from the $\phi\to\eta\gamma$ radiative decays. The branching fraction of the $\eta\rightarrow\pi^0\gamma\gamma$ process is measured to be $\BB(\eta\rightarrow\pi^0\gamma\gamma)=(0.98\pm 0.11_\text{stat}\pm 0.14_\text{syst})\times10^{-4}$. This value is two times smaller than the recent experimental measurements of Refs.~\cite{Prakhov:2008zz,Nefkens:2014zdf} and confirms the preliminary result from KLOE~\cite{KLOE:2005hln}. Our result is in agreement with recent calculations presented in \cite{Escribano:2018cwg} of $\BB=(1.30\pm0.08)\times10^{-4}$. Using the full width of the $\eta$ meson, $\Gamma=(1.31\pm0.05)$ keV \cite{ParticleDataGroup:2024cfk} the partial decay width reads $\Gamma(\eta\to\pi^0\gamma\gamma)=(0.13\pm0.02)$ eV, where the uncertainty is the quadratic sum of the PDG uncertainty value and our measurement.

\begin{figure}[!htbp]
	\begin{center}
		{\includegraphics[height=0.5\linewidth,trim={0 0.2cm 0 1.3cm},clip]{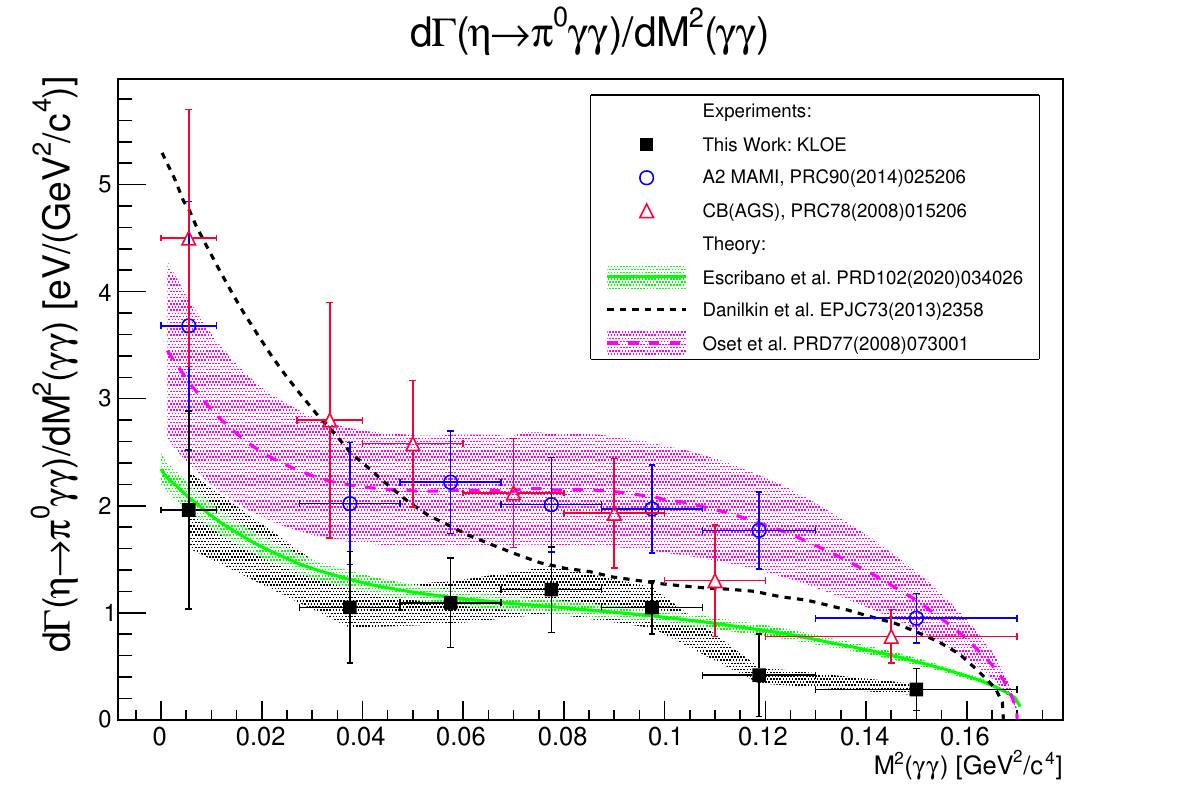}}
		\caption{The measured $d\Gamma(\eta\to\pi^0\gamma\gamma)/dM^2(\gamma\gamma)$ distribution (black points with errors, the grey area represents the common systematics on the data points) compared with previous experiments (red triangles from Ref.~\cite{Prakhov:2008zz} and blue circles from Ref.~\cite{Nefkens:2014zdf}). All errors on $d\Gamma/dM^2$ are total ones. Compared with calculations of Ref.~\cite{Escribano:2018cwg} (green solid line with shaded error band), Ref.~\cite{Danilkin:2012ua} (black dotted line) and Ref.~\cite{Oset:2008hp} (magenta dashed line with shaded error band).}
		\label{fig:dg_plot}
	\end{center}
\end{figure}

The results for  the $d\Gamma(\eta\to\pi^0\gamma\gamma)/dM^2(\gamma\gamma)$ spectrum are given in figure~\ref{fig:dg_plot} and in table~\ref{tab:dgdm2}.
 The systematic uncertainties take into account the contributions listed in table~\ref{tab:syst}. The errors are partially correlated and a part of the uncertainty is common to all data points. This contribution is estimated from the uncertainties of the $\BB$ value and is subtracted from all points. In figure~\ref{fig:dg_plot} our results are compared to the measurements of \cite{Prakhov:2008zz,Nefkens:2014zdf} and to the theoretical predictions from \cite{Oset:2008hp,Danilkin:2012ua,Escribano:2018cwg}.
\begin{table}[h!t!b!p!]
	\centering
	\begin{tabular}{ c |c | c | c | c |c|c|c}
		\hline
		Bin & $1^\text{st}$ & $2^\text{nd}$ & $3^\text{rd}$ & $4^\text{th}$ & $5^\text{th}$ & $6^\text{th}$ & $7^\text{th}$\\ \hline
		Value [eV/(GeV$^2$/c$^4$)]&1.96	&1.05&	1.09	&1.22&	1.05&	0.42&	0.28\\ \hline
		Stat. unc.(\%) & 14.7	&26.1	&26.6	&21.3	&24.5	&57.2	&46.2\\
		Syst.  unc.(\%) &46.4&	43.8&	29.9&	27.7&	10.3&	73.0&	53.7\\
		Total unc.(\%) &48.7&	51.0	&40.0	&35.0&	26.6&	92.7&	70.9\\
		 \hline
	\end{tabular}
	\caption{Values and uncertainties of the $d\Gamma(\eta\to\pi^0\gamma\gamma)/dM^2(\gamma\gamma)$ spectrum. In addition there is an overall normalisation error of 18\%. The bin ranges are $1^\text{st}:(0,0.011)$, $2^\text{nd}:(0.0275,0.0475)$, $3^\text{rd}:(0.0475,0.0675)$, $4^\text{th}:(0.0675,0.0875)$, $5^\text{th}:(0.0875,0.1075)$, $6^\text{th}:(0.1075,0.13)$, $7^\text{th}:(0.13,0.17)$ in GeV$^2$/c$^4$ units.}
	\label{tab:dgdm2}
\end{table}
The value of branching fraction obtained from integration of $d\Gamma/dM^2$ distribution and normalising to the $\eta\to3\pi^0$ channel gives $\BB(\eta\rightarrow\pi^0\gamma\gamma)=(1.00 \pm 0.11_{stat})\times10^{-4}$, which is in good agreement with the value obtained from the full spectrum fit.

\acknowledgments
We warmly thank our former KLOE colleagues for the access to the data collected during the KLOE data taking campaign.
We thank the DA$\Phi$NE team for their efforts in maintaining good running conditions and their collaboration during both the KLOE run
\cite{Gallo:2006yn} and the KLOE-2 data taking with an upgraded collision scheme \cite{Zobov:2010zza, Milardi:2011mz}.
We are very grateful to our colleague G. Capon for his enlightening comments and suggestions about the manuscript.
We want to thank our technical staff: 
M. Anelli for his continuous attention to the gas system and detector safety; 
A. Balla, M. Gatta, G. Corradi and G. Papalino for electronics maintenance; 
C. Piscitelli for his help during major maintenance periods. 
We sincerely thank the technical staff of the LNF computing and web service,
S. Angius, R. Orr\`u, D. Spigone, and M. Tota,
for their support in the maintenance of the KLOE Computing Center.
We would like to express our gratitude to
M. Pistoni for his help and technical support on computing,
L. dell'Agnello, D. Cesini, and C. Pellegrino for their support in the KLOE/KLOE-2 data transfer at CNAF,
A.~Antonelli and M.~Palutan, for their support while heading the LNF research division.

This work was supported in part by the Polish National Science Centre through the Grant No.\
2017/26/M/ST2/00697 and the “Leverhulme Trust LIP-2021-014”.

\bibliographystyle{JHEP} 
\bibliography{lit}

\providecommand{\href}[2]{#2}\begingroup\raggedright\begin{thebibliography}{10}

\bibitem{Holstein:2001bt}
B.R.~Holstein, \emph{{Allowed $\eta$ decay modes and chiral symmetry}},
  \href{https://doi.org/10.1238/Physica.Topical.099a00055}{\emph{Phys. Scripta
  T} {\bfseries 99} (2002) 55}
  [\href{https://arxiv.org/abs/hep-ph/0112150}{{\ttfamily hep-ph/0112150}}].

\bibitem{Oset:2002sh}
E.~Oset, J.R.~Pelaez and L.~Roca, \emph{{$\eta\to\pi^0\gamma\gamma$ decay
  within a chiral unitary approach}},
  \href{https://doi.org/10.1103/PhysRevD.67.073013}{\emph{Phys. Rev. D}
  {\bfseries 67} (2003) 073013}
  [\href{https://arxiv.org/abs/hep-ph/0210282}{{\ttfamily hep-ph/0210282}}].

\bibitem{Danilkin:2012ua}
I.V.~Danilkin, M.F.M.~Lutz, S.~Leupold and C.~Terschlusen, \emph{{Photon-fusion
  reactions from the chiral Lagrangian with dynamical light vector mesons}},
  \href{https://doi.org/10.1140/epjc/s10052-013-2358-1}{\emph{Eur. Phys. J. C}
  {\bfseries 73} (2013) 2358}
  [\href{https://arxiv.org/abs/1211.1503}{{\ttfamily 1211.1503}}].

\bibitem{Oset:2008hp}
E.~Oset, J.R.~Pelaez and L.~Roca, \emph{{$\eta\to\pi^0\gamma\gamma$ decay
  within a chiral unitary approach revisited}},
  \href{https://doi.org/10.1103/PhysRevD.77.073001}{\emph{Phys. Rev. D}
  {\bfseries 77} (2008) 073001}
  [\href{https://arxiv.org/abs/0801.2633}{{\ttfamily 0801.2633}}].

\bibitem{ParticleDataGroup:2024cfk}
{\scshape Particle Data Group} collaboration, \emph{{Review of particle
  physics}}, \href{https://doi.org/10.1103/PhysRevD.110.030001}{\emph{Phys.
  Rev. D} {\bfseries 110} (2024) 030001}.

\bibitem{Escribano:2018cwg}
R.~Escribano, S.~Gonz\`alez-Sol\'\i{}s, R.~Jora and E.~Royo, \emph{{Theoretical
  analysis of the doubly radiative decays $\eta^{(\prime)}\to\pi^0\gamma\gamma$
  and $\eta^\prime\to\eta\gamma\gamma$}},
  \href{https://doi.org/10.1103/PhysRevD.102.034026}{\emph{Phys. Rev. D}
  {\bfseries 102} (2020) 034026}
  [\href{https://arxiv.org/abs/1812.08454}{{\ttfamily 1812.08454}}].

\bibitem{Prakhov:2008zz}
S.~Prakhov et~al., \emph{{Measurement of the invariant-mass spectrum for the
  two photons from the $\eta\to\pi^0\gamma\gamma$ decay}},
  \href{https://doi.org/10.1103/PhysRevC.78.015206}{\emph{Phys. Rev. C}
  {\bfseries 78} (2008) 015206}.

\bibitem{Nefkens:2014zdf}
B.M.K.~Nefkens et~al., \emph{{New measurement of the rare decay $\eta \to
  \pi^0\gamma\gamma$ with the Crystal Ball/TAPS detectors at the Mainz
  Microtron}}, \href{https://doi.org/10.1103/PhysRevC.90.025206}{\emph{Phys.
  Rev. C} {\bfseries 90} (2014) 025206}
  [\href{https://arxiv.org/abs/1405.4904}{{\ttfamily 1405.4904}}].

\bibitem{KLOE:2005hln}
{\scshape KLOE} collaboration, \emph{{The $\eta \to \pi^0 \gamma \gamma$, $\eta
  / \eta^\prime$ mixing angle and the $\eta$ mass measurement at KLOE}},
  {\emph{Acta Phys. Slov.} {\bfseries 56} (2006) 403}.

\bibitem{Zobov:2007xw}
{\scshape DAFNE} collaboration, \emph{{DAFNE status and upgrade plans}},
  \href{https://doi.org/10.1134/S1547477108070042}{\emph{Phys. Part. Nucl.
  Lett.} {\bfseries 5} (2008) 560}
  [\href{https://arxiv.org/abs/0709.3696}{{\ttfamily 0709.3696}}].

\bibitem{Bossi:2008aa}
{\scshape KLOE} collaboration, \emph{{Precision Kaon and Hadron Physics with
  KLOE}}, \href{https://doi.org/10.1393/ncr/i2008-10037-9}{\emph{Riv. Nuovo
  Cim.} {\bfseries 31} (2008) 531}
  [\href{https://arxiv.org/abs/0811.1929}{{\ttfamily 0811.1929}}].

\bibitem{Adinolfi:2002uk}
M.~Adinolfi et~al., \emph{{The tracking detector of the KLOE experiment}},
  \href{https://doi.org/10.1016/S0168-9002(02)00514-4}{\emph{Nucl. Instrum.
  Meth. A} {\bfseries 488} (2002) 51}.

\bibitem{Adinolfi:2002zx}
M.~Adinolfi et~al., \emph{{The KLOE electromagnetic calorimeter}},
  \href{https://doi.org/10.1016/S0168-9002(01)01502-9}{\emph{Nucl. Instrum.
  Meth. A} {\bfseries 482} (2002) 364}.

\bibitem{Adinolfi:2002mvh}
M.~Adinolfi et~al., \emph{{The trigger system of the KLOE experiment}},
  \href{https://doi.org/10.1016/S0168-9002(02)01313-X}{\emph{Nucl. Instrum.
  Meth. A} {\bfseries 492} (2002) 134}.

\bibitem{Ambrosino:2004qx}
F.~Ambrosino et~al., \emph{{Data handling, reconstruction, and simulation for
  the KLOE experiment}},
  \href{https://doi.org/10.1016/j.nima.2004.06.155}{\emph{Nucl. Instrum. Meth.
  A} {\bfseries 534} (2004) 403}
  [\href{https://arxiv.org/abs/physics/0404100}{{\ttfamily physics/0404100}}].

\bibitem{KLOE:2002kzf}
{\scshape KLOE} collaboration, \emph{{Study of the decay $\phi \to \eta \pi^0
  \gamma$ with the KLOE detector}},
  \href{https://doi.org/10.1016/S0370-2693(02)01821-X}{\emph{Phys. Lett. B}
  {\bfseries 536} (2002) 209}
  [\href{https://arxiv.org/abs/hep-ex/0204012}{{\ttfamily hep-ex/0204012}}].

\bibitem{KLOE:2009ehb}
{\scshape KLOE} collaboration, \emph{{Study of the a(0)(980) meson via the
  radiative decay $\phi \to \eta \pi_0 \gamma$ with the KLOE detector}},
  \href{https://doi.org/10.1016/j.physletb.2009.09.022}{\emph{Phys. Lett. B}
  {\bfseries 681} (2009) 5} [\href{https://arxiv.org/abs/0904.2539}{{\ttfamily
  0904.2539}}].

\bibitem{KLOE:2002deh}
{\scshape KLOE} collaboration, \emph{{Study of the decay $\phi \to \pi^0 \pi^0
  \gamma$ with the KLOE detector}},
  \href{https://doi.org/10.1016/S0370-2693(02)01838-5}{\emph{Phys. Lett. B}
  {\bfseries 537} (2002) 21}
  [\href{https://arxiv.org/abs/hep-ex/0204013}{{\ttfamily hep-ex/0204013}}].

\bibitem{KLOE:2006vmv}
{\scshape KLOE} collaboration, \emph{{Dalitz plot analysis of $e^+ e^- \to
  \pi^0 \pi^0 \gamma$ events at $\sqrt{s}$ approximately M($\phi$) with the
  KLOE detector}},
  \href{https://doi.org/10.1140/epjc/s10052-006-0157-7}{\emph{Eur. Phys. J. C}
  {\bfseries 49} (2007) 473}
  [\href{https://arxiv.org/abs/hep-ex/0609009}{{\ttfamily hep-ex/0609009}}].

\bibitem{KLOE:2008woc}
{\scshape KLOE} collaboration, \emph{{Study of the process $e^+ e^- \to \omega
  \pi^0$ in the $\phi$-meson mass region with the KLOE detector}},
  \href{https://doi.org/10.1016/j.physletb.2008.09.056}{\emph{Phys. Lett. B}
  {\bfseries 669} (2008) 223}
  [\href{https://arxiv.org/abs/0807.4909}{{\ttfamily 0807.4909}}].

\bibitem{Hocker:2007ht}
A.~Hocker et~al., \emph{{TMVA - Toolkit for Multivariate Data Analysis}},
  \href{https://arxiv.org/abs/physics/0703039}{{\ttfamily physics/0703039}}.

\bibitem{Voss:2007jxm}
H.~Voss, A.~Hocker, J.~Stelzer and F.~Tegenfeldt, \emph{{TMVA, the Toolkit for
  Multivariate Data Analysis with ROOT}},
  \href{https://doi.org/10.22323/1.050.0040}{\emph{PoS} {\bfseries ACAT} (2007)
  040}.

\bibitem{Barlow:2002yb}
R.~Barlow, \emph{{Systematic errors: Facts and fictions}},  in
  \emph{{Conference on Advanced Statistical Techniques in Particle Physics}},
  pp.~134--144, 7, 2002 [\href{https://arxiv.org/abs/hep-ex/0207026}{{\ttfamily
  hep-ex/0207026}}].

\bibitem{Gallo:2006yn}
A.~Gallo et~al., \emph{{DAFNE status report}}, {\emph{Conf. Proc. C} {\bfseries
  060626} (2006) 604}.

\bibitem{Zobov:2010zza}
M.~Zobov et~al., \emph{{Test of crab-waist collisions at DAFNE Phi factory}},
  \href{https://doi.org/10.1103/PhysRevLett.104.174801}{\emph{Phys. Rev. Lett.}
  {\bfseries 104} (2010) 174801}.

\bibitem{Milardi:2011mz}
C.~Milardi, M.A.~Preger, P.~Raimondi and F.~Sgamma, \emph{{High luminosity
  interaction region design for collisions inside high field detector
  solenoid}}, \href{https://doi.org/10.1088/1748-0221/7/03/T03002}{\emph{JINST}
  {\bfseries 7} (2012) T03002}
  [\href{https://arxiv.org/abs/1110.3212}{{\ttfamily 1110.3212}}].

\end{thebibliography}\endgroup
\end{document}